\documentclass[aps, pra, reprint, longbibliography,
superscriptaddress]{revtex4-1}
\usepackage{hyperref}
\usepackage{graphicx}
\usepackage[utf8]{inputenc}
\usepackage{xcolor}
\usepackage{amsmath, amssymb}
\usepackage{physics}
\usepackage{ulem}
\usepackage{epstopdf}

\begin{document}

\preprint{APS/123-QED}

\title{Finite-temperature phase diagram of dipolar bosons in a cylindrical setup
}

\newcommand{\CFM}[0]{
Centro de F\'{\i}sica de Materiales CFM/MPC (CSIC-UPV/EHU), Paseo Manuel de Lardizabal 5, 20018 Donostia-San Sebasti\'an, Spain}

\author{J. S\'anchez-Baena}
\email{juan.sanchez.baena@upc.edu}
\affiliation{Departament de F\'isica, Universitat Polit\`ecnica de Catalunya, Campus Nord B4-B5, 08034 Barcelona, Spain}

\author{R. Bombín}
\affiliation{\CFM}

\author{J. Boronat}
\affiliation{Departament de F\'isica, Universitat Polit\`ecnica de Catalunya, Campus Nord B4-B5, 08034 Barcelona, Spain}

\date{\today}

\begin{abstract}
 Temperature plays a non-trivial role in dipolar systems, 
being able to either induce or destroy spatial density modulations.
In order to shed light into the interplay between thermal effects and 
quantum correlations, we employ the path integral Monte 
Carlo method (PIMC) to obtain the phase diagram of a system of dipolar bosons in
an infinite cylindrical geometry and at finite temperatures. For a fixed
total density, we vary the scattering length of the interaction and the 
temperature, and characterize the structure and superfluidity of the emerging 
phases. The PIMC method accounts for thermal effects and quantum correlations at 
all orders, and can be used to quantitatively estimate the superfluid fraction 
of the system. Remarkably, we 
find a regime of parameters where, starting from a supersolid, 
the spatial order vanishes with temperature 
 before the superfluidity does. We also find 
a region of the phase diagram where the superfluid fraction displays a 
non-monotonous dependence with temperature. Our results extend previous 
descriptions, relying on Bogoliubov theory, to larger temperatures in the thermodynamic limit and 
illustrate how spatial modulations and superfluidity are lost as temperature is
increased.
\end{abstract}

\maketitle

\section{\label{sec:introduction}Introduction}

The field of ultracold dipolar atoms has attracted a considerable theoretical 
and experimental interest in the recent years~\cite{chomaz:rpp2023}. The combination of anisotropy and
long-range properties of the dipole-dipole interaction (DDI) gives rise to a 
rich variety of physical phenomenal, the main highlight being a dipolar 
supersolid of droplets: an exotic quantum state that combines superfluidity 
with 
diagonal long-range order. Dipolar supersolids have been 
extensively studied theoretically and realized 
experimentally~\cite{stuhler2005:prl,Wenzel:2017ch,ferlaino16,Tanzi:Nature:2019, 
Pfau:PRX:2019,Ferlaino:PRX:2019,Natale:PRL:2019,Modugno:PRL:2019, 
Tanzi:Science:2021,HertkornPRX21,
Ferlaino:PRL:2021,norcia21:nature,Biagioni:PRX:2022,he2025:scienceadv,
he2025:arxiv}. Most of the previous theoretical works on dipolar systems have 
focused in the zero temperature 
regime~\cite{santos2003:prl,sinha2007:prl,pelster12,Baillie_2015,wachtler16:PRA,
bombin17:PRL,baillie2017:PRL,edler2017:prl,zhang19,blackie20:prr,zhang21:pra,
hertkorn21,guijarro22:prl,staudinger23:PRA,baena23shell,ripley2023:pra,
guijarro2024,aleksandrova_2024_PRA,blakie2025:prl,baena2025:pra}, while some others
have accounted for thermal effects making use of the the random phase 
approximation~\cite{pengtao2021}, c-field 
theories~\cite{pawlowski2013:PRA,bland22:prl,proukakis2025:arxiv}, the complex 
Langevin algorithm~\cite{heinen2025:arxiv}, Bogoliubov 
theory~\cite{ticknor2012:PRA,ticknor2012:PRA2,bisset12:pra,boudjemaa15:PRA,
oktel19,oktel20,baena22,baena24,He2024:PRR,baena26:arxiv}, diagrammatic 
theories~\cite{Glaum:PRL:2007,Glaum:PRA:2007} or Monte Carlo 
methods~\cite{bombin19:PRA,shi25:prx,bombin25:prl}. It is specifically relevant 
to consider finite temperature in the description of dipolar systems.  
In fact, in the strongly dipolar regime ($a_{\rm dd} > a$, where $a_{\rm dd}$ 
is the dipole length and $a$ is the scattering length) it has been 
experimentally shown that, strikingly, thermal fluctuations can have a huge 
qualitative effect on the system, promoting spatial 
modulations~\cite{Ferlaino:PRL:2021}. These puzzling results were subsequently 
reproduced and explained using Bogoliubov theory~\cite{baena22}, which showed 
that thermal atoms compress the 
cloud, enhancing the role of the DDI and leading to the emergence of modulated 
states.

Up to now, the picture of the interplay between temperature, spatial order, and 
superfluidity in dipolar systems remains incomplete. In strictly two-dimensions 
(2D), results obtained with the random phase approximation (RPA) show that 
thermal effects act classically, melting the solid~\cite{pengtao2021}. Quantum 
Monte Carlo in 2D reported results of the superfluidity of  
dipolar stripes in the highly correlated regime, and showed that the loss of 
superfluidity precedes that of spatial modulations as temperature is 
increased~\cite{bombin19:PRA}. Crucially, both these works focus on the 
thermodynamic limit in two dimensions, where thermal effects prevent the 
existence of a Bose-Einstein Condensate (BEC) at any finite temperature. 
Instead, Ref.~\cite{baena26:arxiv} studies a trapped system in the quasi-2D 
regime featuring a finite value for the condensate fraction.
By using Bogoliubov theory, it is shown that thermal excitations can 
promote a modulational instability. Moving to three dimensions, the system considered here has been studied using
Bogoliubov theory~\cite{baena22,baena24,He2024:PRR}, which is only valid in the
limit of large condensate fractions (or low temperatures). On the other hand, a 
trapped setup in a cigar-shape geometry has been studied with the stochastic
Gross-Pitaevskii equation~\cite{bland22:prl}, which constitutes an approximate 
high temperature theory, and the path integral Monte Carlo (PIMC) 
method~\cite{bombin25:prl}. The latter work accounts exactly for thermal and 
quantum fluctuations in a wide temperature range, bridging the phenomenology of 
induction with that of destruction of modulations by heating. However, due to 
the small, finite-size system considered, the results do not allow to elaborate 
a proper phase diagram in the thermodynamic limit, from which more general 
conclusions could be drawn. Moreover, an extensive exploration of the
variation of scattering length of the interaction together with
temperature is also absent.

\begin{figure}[t]
\centering
\includegraphics[width=\linewidth]{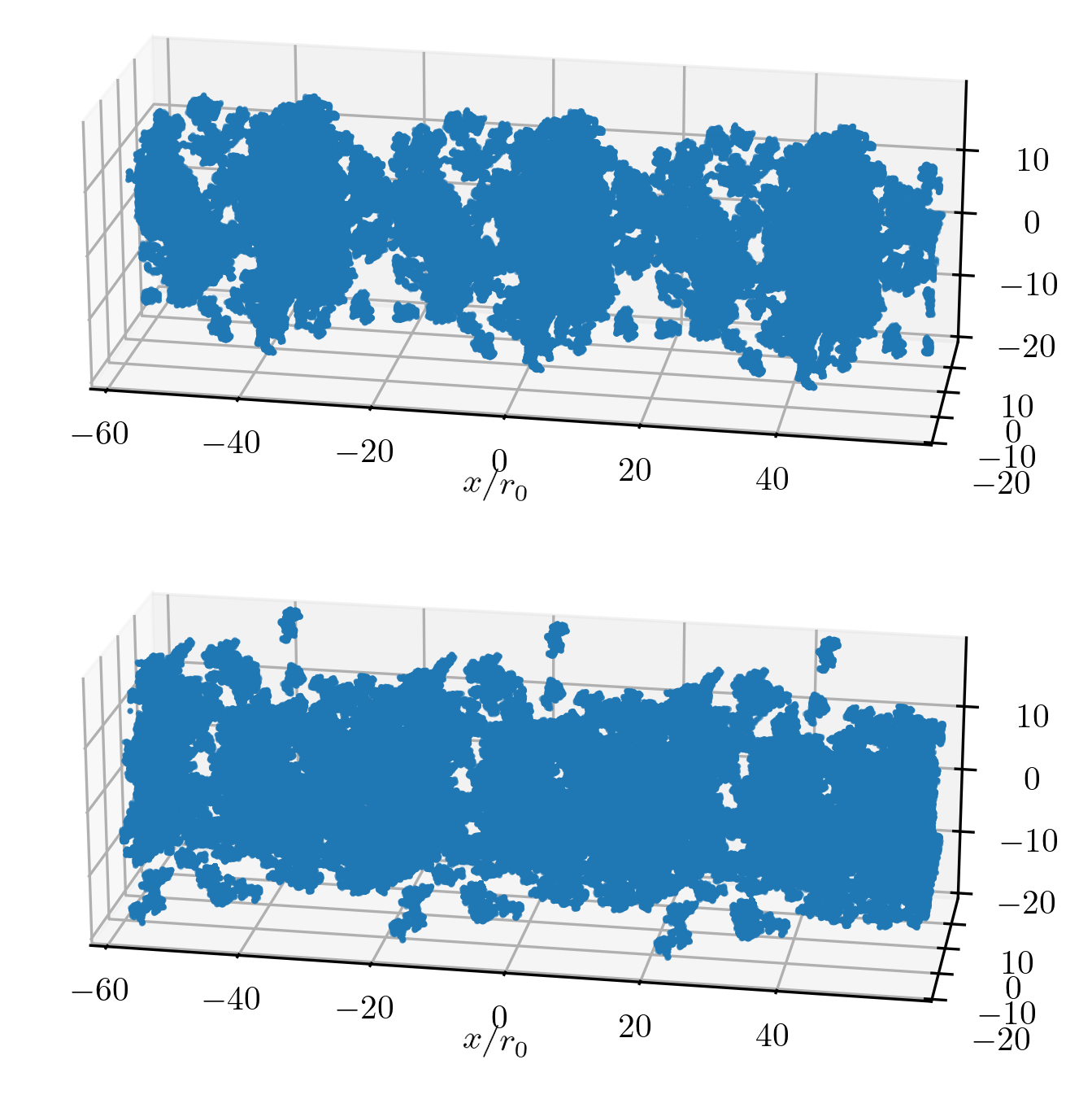}
\caption{Representative snapshots of the supersolid (top) and superfluid 
(bottom) phases of the dipolar system confined in an infinite tube. The 
temperature is set to $T=572$ nK while the scattering length values are $47.31$ 
$a_0$ (top) and $56.74$ $a_0$ (bottom). The simulation box of length $L_x = 40 
r_0$ has been replicated to produced the plots.}
\label{fig0}
\end{figure}

In the present work, we fill part of this void by studying a system 
of dipolar bosons in an infinite cylinder and at finite temperature. In this
geometry, the system is expected to host a transition between an unmodulated BEC 
and a solid one-dimensional (1D) array of dipolar clusters made up of several 
atoms, with a supersolid phase existing in between 
~\cite{Roccuzzo2019,blackie20:prr,smith22,Buchler:PRA:2022,baena22,baena24,
Schubert2026}. In this case, at temperatures below the transition to the normal 
phase, the gas unbounded to the clusters facilitates the necessary particle 
exchange for the rise of superfluid flow along the infinite axis. Examples of 
supersolid and superfluid states are shown in Fig.~\ref{fig0}. We employ the 
PIMC method and perform \textit{ab-initio} calculations that take into account 
both quantum correlations and thermal fluctuations to all orders. For a fixed 
total density, we vary the temperature and the scattering length and 
characterize the structure and the superfluidity of the emerging quantum states. 
With this information, we elaborate a phase diagram in the range of temperatures 
$T \in (0.5 T_c, 1.1 T_c)$, with $T_c$ the critical temperature at which the system 
transitions to the normal fluid phase. For a system of $^{162}$Dy atoms and the geometry considered here, 
this equals to $T_c \simeq 1.11$ $\mu$K. This
allows us to specifically address how superfluidity and spatial order are lost 
as temperature is increased. 

The paper is organized as follows. In Section~\ref{sec:system} we introduce the dipolar system confined in the infinite tube geometry. Computational details about the PIMC method are given in Sec.~\ref{sec:method}. In Section~\ref{sec:results} we discuss our results and, finally in Section~\ref{sec:conclusions} we summarize some conclusions.

\section{\label{sec:system}System and Hamiltonian}

We consider a system of $N=300$ $^{162}$Dy dipolar atoms of mass $m$ in a tubular geometry with
trapping frequencies $(\omega_x,\omega_y,\omega_z) = 2 \pi \times (0, 5050, 
5050)$ Hz and periodic boundary conditions (PBC) along the $x$-axis. The 
Hamiltonian of the system is
\begin{equation}
H=-\frac{\hbar^2}{2m}\sum_{i=1}^{N} \nabla_i^2 + \sum_{i<j}^{N} V({\bf 
r}_i-{\bf 
r}_j) \ ,
\label{hamiltonian}
\end{equation}
with ${\bf r}_i$ the coordinate of the ith particle and the inter-particle interaction modeled as
\begin{equation}
 V({\bf r}) = \frac{C_{12}}{r^{12}} - \frac{C_6}{r^6} +
 \frac{C_{\rm dd}}{4 \pi} \frac{ \left( 1 - 3 \cos^2 \theta  \right) }{r^3}
 \ ,
 \label{potential}
\end{equation}
which corresponds to a fully polarized system along the $z$-axis, with
${\bf r} = {\bf r}_i - {\bf r}_j$, $r = \abs{{\bf r}}$, and $\theta$ the angle between the vector ${\bf r}$ and the $z$ axis.
In the following, we use reduced units, with the
characteristic length and energy scales defined by
$r_0 = m C_{\rm dd}/(4\pi \hbar^2) = 3 a_{\rm dd}$ and $\epsilon = \hbar^2/(m r_0^2)$, where $a_{\rm dd} = 129.2 a_0$ for $^{162}$Dy atoms and $a_0$ is the Bohr radius. In
these units, $C_6 = 0.02978$~\cite{Li2017}, while
the value of $C_{12}$ is determined to reproduce a given $s$-wave scattering 
length $a_s$ through a T-matrix 
calculation~\cite{bottcher19}.

The length along the $x$-axis of the simulation box is fixed to $L_x = 40 r_0$ 
which, together with the oscillator lengths in the trapped directions $l_{\xi} = 
\sqrt{\frac{\hbar}{m \omega_{\xi}}}$, $\xi = y,z$,  yields a mean density that 
we roughly estimate as $\rho_{\rm mean}= 1.8 \times 
10^{15}$ cm$^{-3}$. This choice of parameters results into a value of the 1D 
density of $n_{\rm 1D} = 365$ $\mu$m$^{-1}$, which is fixed for all the 
calculations in this work.
The mean density is slightly larger than the one of 
Refs.~\cite{HertkornPRX21,bombin25:prl}, which consider a fully trapped system 
elongated along the $x$-axis, but with a $N/(l_y l_z)$ ratio is similar to our 
configuration. 
Our choice for $L_x$ is motivated by the PIMC results of 
Ref.~\cite{bombin25:prl}, which show a two-droplet supersolid with a typical 
separation between droplets of $d \simeq 20 r_0$. In principle, for every value 
of the scattering length and temperature, one 
should optimize $L_x$ by minimizing the free energy. However, the large 
computational cost of PIMC simulations makes this prohibitive. Nevertheless, to 
benchmark the approximation of keeping the box size fixed, we run extended 
Gross-Pitaevskii simulations (eGPE)~\cite{wachtler16:PRA} incorporating the 
effect of quantum fluctuations~\cite{Lima:2011eq,Lima:2012hr} at zero 
temperature and study the BEC-supersolid transition as a function of the 
scattering length. We do so following two approaches: fixing $L_x = 40 r_0$ as 
it is done in the PIMC calculations and optimizing the energy per particle as a 
function of $L_x$. Clearly, the particle number in the eGPE simulations is rescaled together with $L_x$, such that the 1D density always matches the value considered in the PIMC calculations.
We calculate the contrast of the column density, defined as
\begin{equation}
C = 
\frac{\rho_{\rm max} - \rho_{\rm min}}{\rho_{\rm max} + \rho_{\rm min}} \ ,
\label{contrast}
\end{equation}
where 
$\rho(x) = \int dy dz \abs{\Psi(x,y,z)}^2$. The modulational transition takes place
when $C$ becomes zero as the scattering length is increased. We find the values 
of the scattering length in the transition from the eGPE calculations to be $a = 42.37$ $a_0$ for $L_x = 40
r_0$ and $a = 44.18$ $a_0$ for the optimized box length $L_{\rm opt.} = 29$ $r_0$. The small shift in the
scattering length of ($\Delta a \simeq 1.8 a_0$) is below the standard 
resolution in experiments~\cite{Ferlaino:PRL:2021,he2025:arxiv}, which validates 
our approximation.
\begin{figure}[t]
\centering
\includegraphics[width=0.9\linewidth]{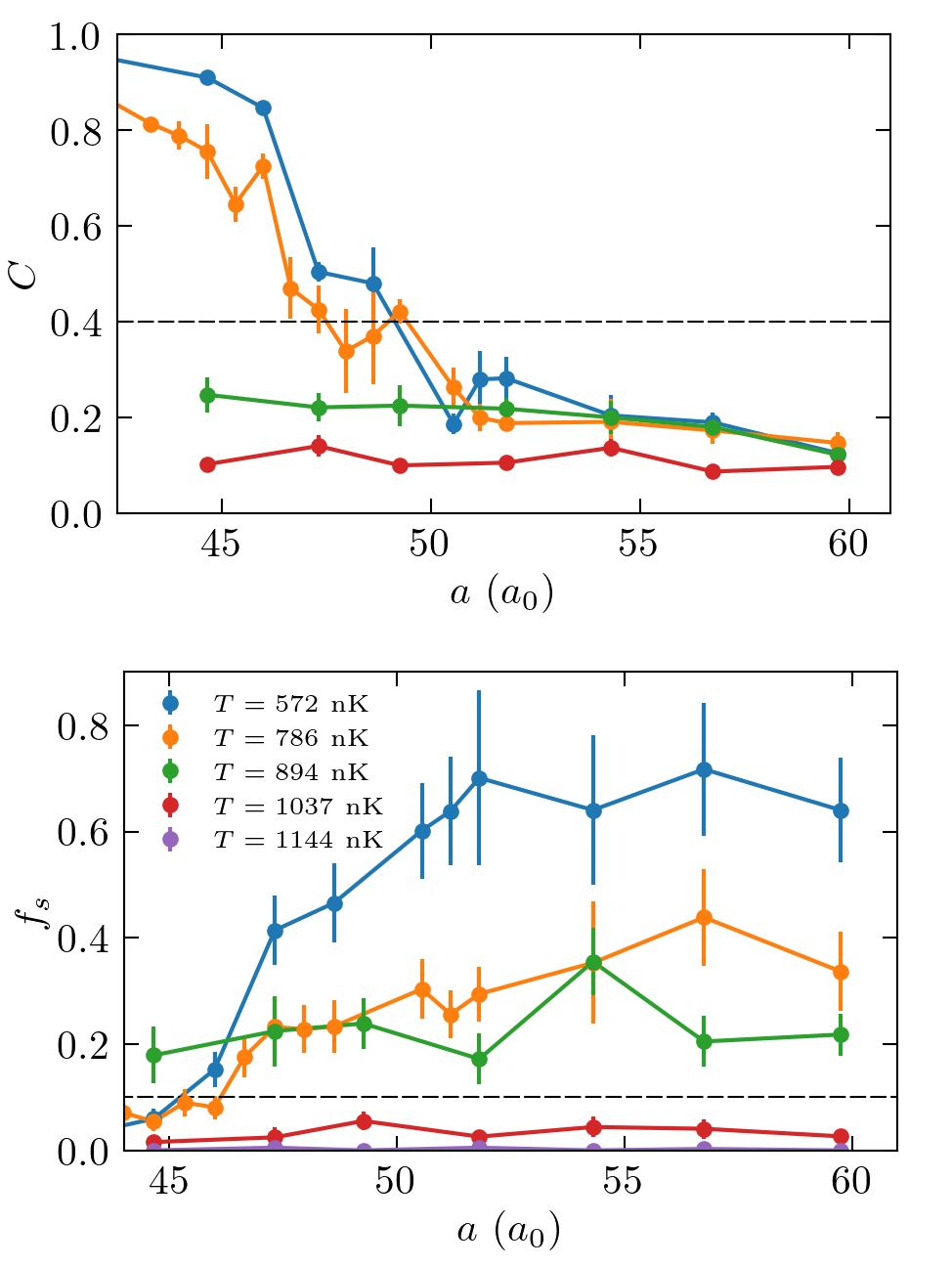}
\caption{Top: contrast of the $x$-column density (see the text) as a function of 
the scattering length for different values of the temperature.
Bottom: superfluid fraction as a function of the
scattering length for different values of the temperature. The dashed lines 
indicate the thresholds of $C$ and $f_{\rm s}$ for the separation between the 
different phases (see the text and Table~\ref{tab:table1}). The color code for the temperature is the same in both plots.}
\label{fig2}
\end{figure}

\section{\label{sec:method}The Path Integral Monte Carlo method}

To study the aforementioned system, we use the path integral Monte Carlo method.
For a given Hamiltonian, the PIMC method provides exact results (up to some statistical uncertainty) 
for the energy, structure, and superfluidity of a quantum state at a given 
temperature $k_{B}T$,
with $k_{B}$ the Boltzmann constant.
Due to its quantitative predictive power, the method has been widely used in the past to study, for example, the phase diagram and 
superfluid transition in $^4$He~\cite{Ceperley1989,Ceperley1995},
systems of magnetic atoms~\cite{Filinov2010,bombin19:PRA,bombin25:prl} as is the case here, 
and, more recently microwave-shielded electric dipolar molecules~\cite{shi25:prx,ciardi25:prl,ciardi26:arxiv}.
The zero temperature version of PIMC has been used to demonstrate the phase coherence between droplets in supersolid arrays~\cite{Bombin24:NJP}.

At any temperature, and mainly at $T$ below the critical temperature $T_c$,
 the PIMC calculation requires an accurate action to reduce 
the number of imaginary-time steps (\textit{beads}) representing each atom 
(\textit{polymer}) to a manageable level.
To this end,  we use the fourth order 
Chin's action~\cite{Takahashi1984,Takahashi1984a,Chin2004,Chin2002}, that can 
be made to work effectively up to sixth order for the energy estimation by 
optimizing its control  parameters~\cite{Sakkos2009}.
Efficiency in the sampling of permutations between identical particles is 
also fundamental to obtain accurate results for the one-body density matrix 
and superfluid densities. To get it right we use the Worm Algorithm, that has 
proven its accuracy in different systems~\cite{Boninsegni2006}.

In order to quantitatively characterize the superfluid to normal phase transition in the system, we 
need to evaluate the superfluid density. In the PIMC method, this is done 
through the well known winding number estimator~\cite{Pollock1987},
\begin{equation}
f_{\rm s}=\rho_{\rm s}/\rho = \frac{m k_B T}{N \hbar^2}\langle \textbf{W}^2\rangle \ ,
\label{eqn:Winding}
\end{equation}
where \textbf{W} is the  so called winding number, estimated along the $x$ axis 
where periodic boundary conditions apply. 

\section{\label{sec:results}Results}

Our main goal is to calculate the phase diagram for the dipolar system confined 
in the infinite tube in terms of the temperature and the scattering length. As 
mentioned previously, the system can be either in the solid, supersolid, 
superfluid and normal fluid phases. Each phase can be labeled based on its 
structural and superfluid properties, which we quantify with the contrast of the 
column density along the $x$-axis
and the superfluid fraction $f_{\rm s}=\rho_{\rm s}/\rho$ along the same axis. 
Fig.~\ref{fig2} depicts the values of $C$ and $f_{\rm s}$ as a function of the 
scattering length for different temperatures in the range $T\in[0.5,1.1]T_{\rm 
c}$.
Looking at the contrast, we can see that, at  low temperatures ($T<800$~nK), it 
decreases as the scattering length is increased, showing a steep decline 
followed by a plateau with some peaks reaching up to $C \simeq 0.3 \pm 0.05$. 
This is due to the statistical fluctuations in a single PIMC simulation, which give rise to a small finite contrast even for the case of an homogeneous phase. In view of this, for practical reasons, we set a threshold of $C=0.4$ to separate the modulated and fluid phases.
In regards to the superfluid fraction, we can see that it vanishes at the largest temperatures considered, indicating the transition to the normal fluid phase, as expected. 
Besides that, the behavior of $f_{\rm s}$ 
is richer at low temperatures.
For low values of the scattering length, where the contrast is large, the
superfluid fraction never identically vanishes in this range of temperatures, but rather flattens fluctuating 
around small values $0<f_{\rm s}<0.1$. Concordantly, there is always a very small remanent of gas in between the dipolar
clusters in our PIMC simulations, which contributes to the superfluid signal at 
low temperatures, below the transition to the normal phase. Nevertheless, we 
consider this small superfluid signal to be negligible and thus set a threshold 
value of $f_{\rm s} = 0.1$ to distinguish between the solid and supersolid phases.
We summarize our choice for the labeling of the different phases in terms of $C$ and $f_{\rm s}$ in Table~\ref{tab:table1}.

\begin{table}
\begin{center}
\begin{tabular}{ | c | c | c | }
 \hline Phase & $C$ & $f_{\rm s}$ \\ \hline
 Solid (S) & $>0.4$ & $<0.1$ \\ \hline
 Supersolid (SS) & $>0.4$ & $>0.1$ \\ \hline
 Superfluid (SF) & $<0.4$ & $>0$ \\ \hline
 Normal fluid (NF) & $<0.4$ & $0$ \\ \hline
\end{tabular}
 \caption{\label{tab:table1}Labeling of the different phases that arise in the dipolar system confined in an infinite tube at finite temperature in terms of the contrast $C$ (see the text) and the superfluid fraction $f_{\rm s}=\rho_{\rm s}/\rho$. The arguments behind the choice of the thresholds in $C$ and $f_{\rm s}$ is discussed in the text.}
\end{center}
\end{table}

\begin{figure}[t]
\centering
\includegraphics[width=\linewidth]{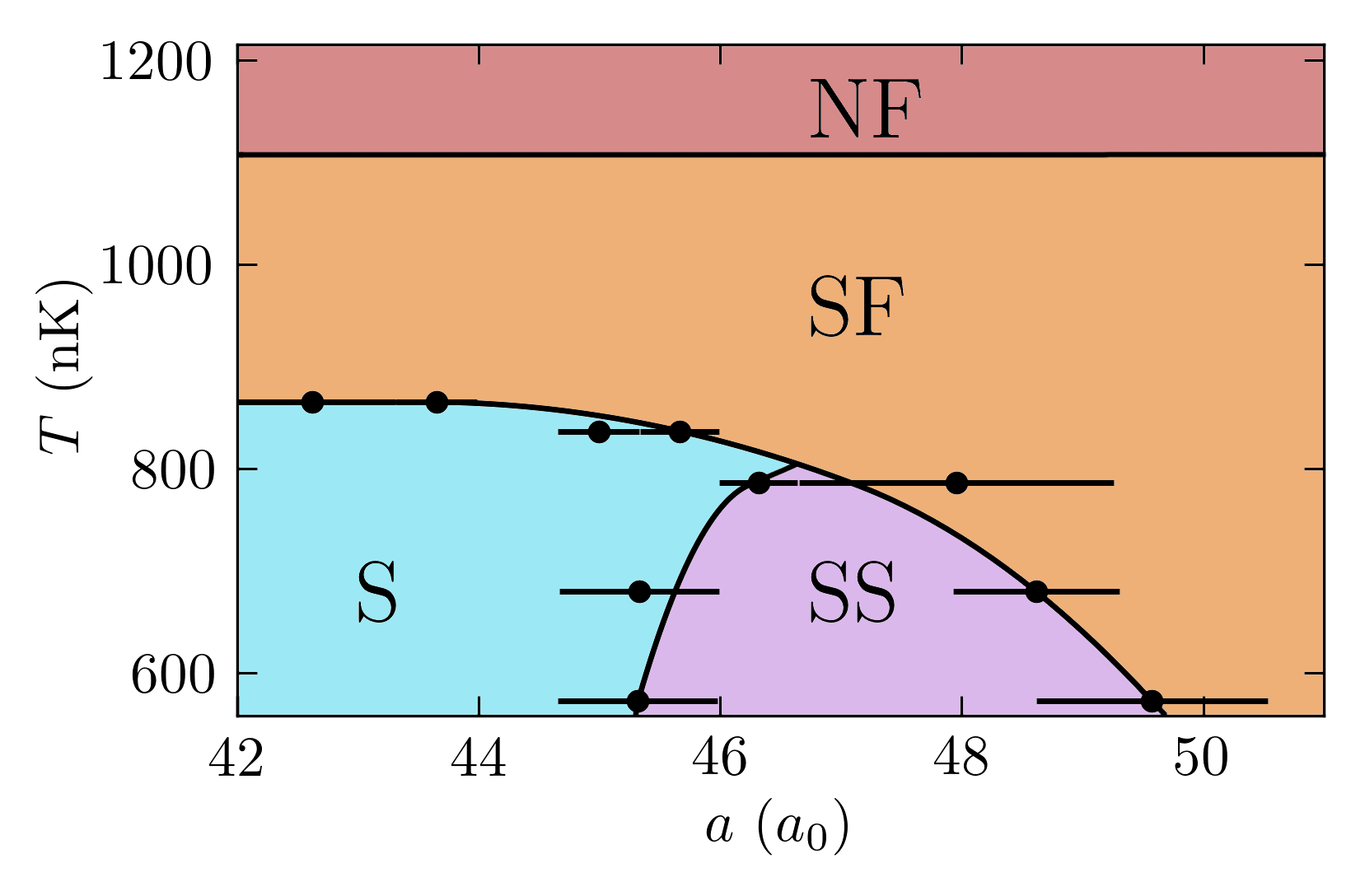}
\caption{Phase diagram for the dipolar system in the infinite tube as a function 
of the scattering length, $a$, and temperature, $T$. The solid (S), supersolid 
(SS), superfluid (SF) and normal fluid (NF) phases are labeled.}
\label{fig3}
\end{figure}

\begin{figure*}[t]
\centering
\includegraphics[width=0.6\linewidth]{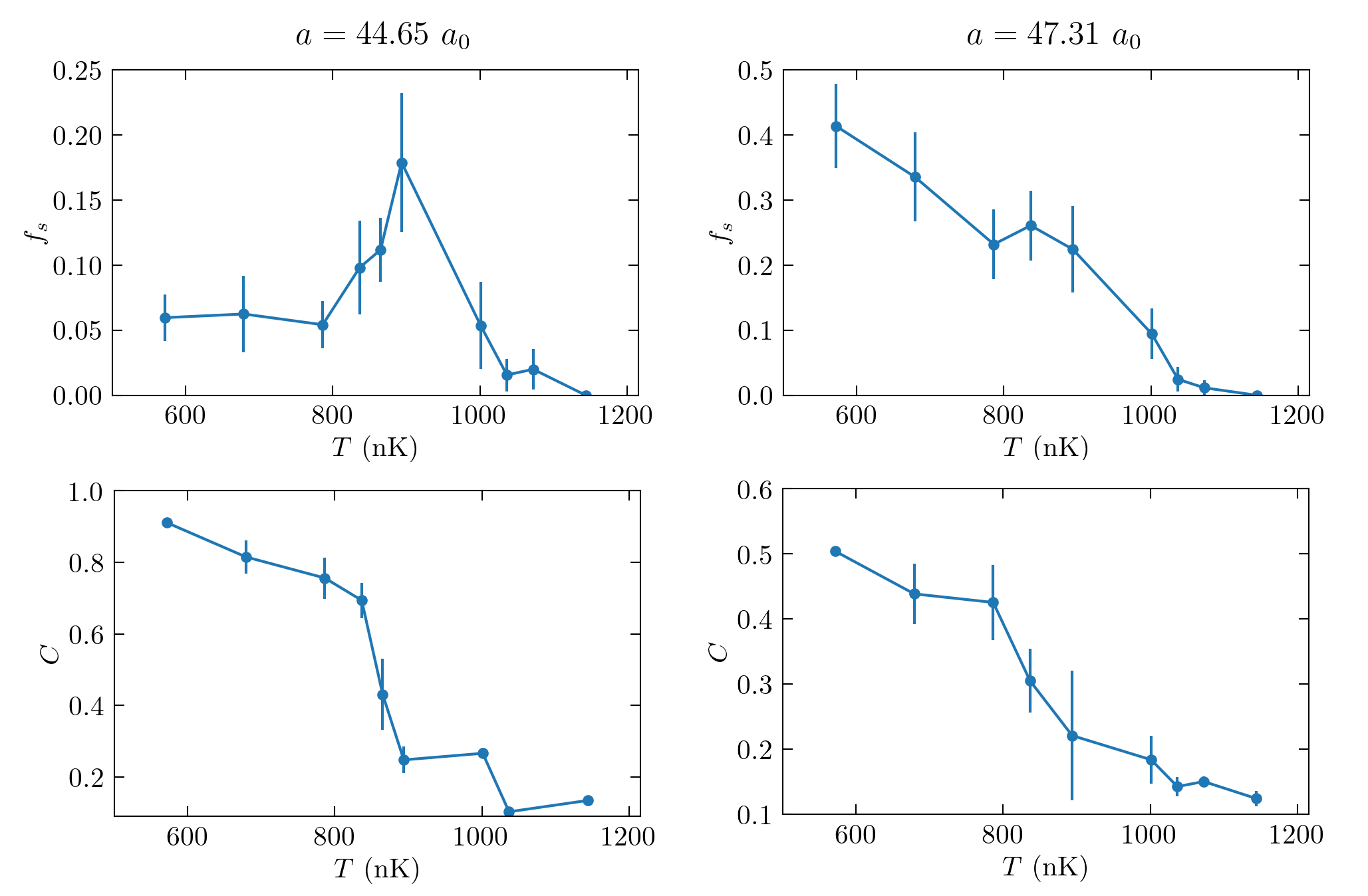}
\caption{Superfluid fraction and contrast of the column density along the $x$-axis as a function of the temperature for a scattering length of  $a = 44.56$ $a_0$ (left) and  $a = 47.31$ $a_0$(right).}
\label{fig4}
\end{figure*}

Previous results for the same geometry relying on mean-field approaches show 
that the modulational transition can be of either first or second order 
depending on the 1D density of the 
system~\cite{blackie20:prr,smith22,baena24,Buchler:PRA:2022}. In this case, the 
results of Fig.~\ref{fig2} show an apparently continuous variation of the 
contrast and superfluid fraction as the transition is crossed. Therefore, for 
our parameters of choice, we can conclude that the transition is of continuous 
nature. This is in agreement with our eGPE calculations at zero temperature.
Nevertheless, despite the solid-supersolid transition has been predicted and 
extensively studied using mean-field approaches, it is discussed in 
Ref.~\cite{smith22} that a quantitatively accurate description of such 
transition, involving an insulating phase, is beyond the scope of applicability 
of the eGPE. Thus, our PIMC calculations allow to more accurately characterize 
this transition at finite temperature and establish a satisfactory comparison 
with the eGPE zero-temperature results. 

Having established the criteria to distinguish between the different phases, we present in Fig.~\ref{fig3} the phase diagram of the system as a function of the scattering length and temperature, which constitutes the main result of our work. 
For low values of the temperature, as the scattering length increases, a transition from solid to supersolid and from that to a superfluid is observed, as expected~\cite{blackie20:prr,smith22,baena24,Buchler:PRA:2022}.
Remarkably, the superfluid signal remains finite well above the melting temperature of the modulated phases.
Starting from the supersolid phase, this implies that, as temperature is 
increased while leaving the total density and the scattering length constant, 
spatial modulations are first destroyed by 
thermal effects, yielding a superfluid. As temperature is further increased up 
to the critical temperature $T_c = 1.1 \pm 0.04$ $\mu$K, the superfluid signal is lost, leading to the normal fluid
phase.  From our simulations we observe that the change in $T_c$ due to the interaction is below a $4\%$ 
for the values of the scattering length considered.
This thermal evolution contrasts with the one reported previously for dipolar 
systems in 
other geometries.
In a 2D system, the bulk dipolar system in the plane features a stripe phase 
for tilted dipoles~\cite{macia14}. In that case, as temperature is increased, 
thermal effects destroy superfluidity first, yielding a striped solid that 
subsequently melts at greater temperatures~\cite{bombin19:PRA}. 
This may be explained by the drastically different regimes of correlation of the two systems: while our system sits in the dilute regime, with a mean density and scattering length of the order of those in current experiments, the system of Ref.~\cite{bombin19:PRA} is extremely dense, with 2D densities of the order of $n_{\rm 2D} r_0^2 \sim 10$, which imply a gas parameter in 2D of $x_{\rm 2D} = n_{\rm 2D} a_{\rm 2D}^2 \sim 27.3$, way above the dilute regime. 
In these conditions, the stripe structure which is entirely induced by the 
anisotropy and strength of the interaction becomes more robust to thermal 
effects. 
This contrasts with the supersolid states that are formed of droplet arrays, which are stabilized by quantum fluctuations at low temperature.
The scenario found in our system also differs from the one arising during the experimental evaporation protocol for a Dysprosium cloud~\cite{Ferlaino:PRL:2021}. 
However, in that experiment, the system is fully trapped and far from the thermodynamic limit, since only three clusters form in the modulated state.  
More importantly, the atom number in that setup dramatically decreases while the system is cooled down and modulations rise, which may favor the formation of modulations at their 1D density of roughly $n_{\rm 1D} \sim 10^4$ $\mu$m$^{-1}$~\cite{blackie20:prr,baena24}. 
Meanwhile, the 1D density is kept constant in our calculations and sits at a considerably lower value of $n_{\rm 1D} = 365$ $\mu$m$^{-1}$, which also explains why the modulational transition takes place at significantly lower values of the scattering length.

A second noticeable feature of the phase diagram of Fig.~\ref{fig3} is the 
presence of a solid-superfluid phase boundary. This implies that, as temperature 
increases, a non-superfluid solid melts into a superfluid as the spatial 
structure disappears. Note, that further increasing the temperature makes the 
superfluidity vanish, as expected. However, this implies a non-monotonous 
behavior of the superfluid fraction with temperature, as we illustrate in 
Fig.~\ref{fig4}, where we report $f_{\rm s}$ and the contrast $C$ as a function of the 
temperature $a = 44.56$ $a_0$. 
This mon-monotonous trend stems from two competing effects on the superfluid fraction of the system: the destruction of spatial modulations tends to increase it while thermal effects reduce it. As the scattering length increases, the system becomes less dipolar, decreasing the contrast of the structures and thus enlarging the superfluid fraction. This restores the expected, monotonous decreasing trend of the superfluid fraction with increasing temperature, as it is also illustrated in Fig.~\ref{fig4} for $a = 47.31$ $a_0$.

\section{\label{sec:conclusions}Conclusions}

In conclusion, we have calculated the phase diagram of a polarized dipolar 
system 
in an infinite tube via the PIMC method, which accounts for 
the effects of quantum and thermal fluctuations to all orders. We focus on the 
relevant temperature range $T \in (0.5 T_c, 1.1T_c)$ where the different phases
are observed. For the parameters considered here, we have a dilute system with a 
mean 3D density not far from the one present in current state of the art
experiments~\cite{Kadau:2016cb}. 
Starting in the supersolid phase at the lowest temperature considered and heating 
the system with constant scattering length and 1D density, the spatial modulations 
are first destroyed, leading to the superfluid phase. Upon further increasing the 
temperature, superfluidity is lost, as expected.
If one instead considers a lower value of 
the scattering length where the system starts in the solid phase, heating first 
destroys modulations, increasing superfluidity and leading the system to the 
superfluid phase, which then transitions to the normal fluid phase as 
temperature further rises. This reflects a non-trivial, non-monotonous behavior 
of the superfluid fraction along the longitudinal axis with respect to 
temperature, which arises from the competition of two effects: the increase of 
superfluidity induced by the destruction of spatial order and its decrease by 
thermal effects.

Despite focusing on the temperature regime close to the critical point, $T \in 
(0.5 T_c, 1.1 T_c)$, the PIMC results at the lowest temperature
considered and the eGPE results at $T=0$ are compatible with a thermal promotion of 
spatial modulations for $T< 0.5 T_c$. 
This is because we find the scattering length values $a_{\rm PIMC} = 49.37 \pm 0.95$ $a_0$ at $T \simeq 0.5 T_c$ and $a_{\rm eGPE}= 44.18$ $a_0$ at $T=0$ for 
the supersolid-superfluid transition. This is consistent with previous 
results~\cite{baena22,baena24,He2024:PRR,bombin25:prl,baena26:arxiv}.

Nevertheless, fully addressing this question will require to perform ab-initio 
calculations at temperatures lower than $0.5 T_c$ or at zero temperature, which 
go beyond the eGPE and may yield significant quantitative corrections to 
mean-field results~\cite{bottcher19}.

Our results offer quantitative insight into the intricate interplay between 
superfluidity, spatial modulation and thermal effects at temperatures close to 
the critical point and in a clean setting in the thermodynamic limit. The 
different relative thermal robustness of spatial modulations and superfluidity 
found in our case with respect to other dipolar systems in different geometries 
and density regimes~\cite{bombin19:PRA,Ferlaino:PRL:2021} suggests that a more 
exhaustive characterization is needed to understand the full picture. In this 
sense, interesting future perspectives include the exploration of BKT physics in 
quasi-2D geometries, where thanks to the tuning of the tilting angle of the 
dipoles, dilute striped supersolid states are 
obtained~\cite{staudinger23:PRA,baena2025:pra,he2025:arxiv}. This lower level of 
inter-atomic 
correlations with respect to the set-up of Ref.~\cite{bombin19:PRA} should 
decrease the robustness of the stripe structure with temperature and potentially 
lead to the same phenomenology presented here. 

\section{\label{sec:acknowledgements}Acknowledgements}

J. S. B. and J. B. acknowledge support by the Spanish Ministerio de Ciencia, Innovación y Universidades (grant PID2023-147469NB-C21, financed by MICIU/AEI/10.13039/501100011033 and FEDER-EU).
\bibliography{paper_dipoles_tube_pimc}

\end{document}